\documentstyle[12pt,twoside,fleqn,espcrc1]{article}

\newcommand{\da}{\; \dagger}

\newcommand{\AmS}{{\protect\the\textfont2
  A\kern-.1667em\lower.5ex\hbox{M}\kern-.125emS}}
\newcommand{\gapproxeq}{\lower.7ex\hbox{$\;\stackrel{\textstyle>}{\sim}\;$}}
\newcommand{\arrowslash}{\lower.1ex\hbox{$\;\stackrel{{\tiny not}}{\rightarrow}
\;$}}

\input{psfig.sty}

\title{ \vspace{-3cm}\begin{flushright} 
\small{hep-ph/0007199}
\\ \small{LA-UR-00-2142}
 \end{flushright} \vspace{.5cm}
Pseudovector mesons, hybrids and glueballs}

\author{Leonid Burakovsky and Philip R. Page\address{Theoretical Division, MS-B283, Los Alamos
National Laboratory, Los Alamos, \\ NM 87545, USA.  }\footnote{{\it E-mail:} burakov@lanl.gov, prp@lanl.gov}\footnote{Talk given by PRP at the 7th Conference 
on the Intersections of Nuclear and Particle Physics (CIPANP 2000), Quebec City,
 Canada, May 22--28, 2000.}}
       
\begin{document}

\maketitle

\begin{abstract}
We consider glueball-- (hybrid) meson mixing for the low--lying four 
pseudovector states. The $h_1^{'}(1380)$ decays dominantly
to $K^{\ast}K$ with some presence in $\rho\pi$ and $\omega\eta$.
The newly observed $h_1(1600)$ has a 
$D$-- to $S$--wave width ratio to $\omega\eta$ which makes its
interpretation as a conventional meson unlikely. 
We predict the decay pattern of the isopartner 
conventional or hybrid meson $b_1(1650)$. A notably 
narrow  $s\bar{s}$ partner
 $h_1^{'}(1810)$ is predicted.
\end{abstract}

\vspace{.5cm}

The pseudovector ($J^{PC}=1^{+-}$) $s\bar{s}$ ground state has the interesting
property that its OZI allowed decay to open strangeness, 
i.e. $K^{\ast}K$, 
which is {\it a priori} expected to be dominant, is severely suppressed by
phase space. This not only makes the state anomalously narrow
\cite{pdg98}, but opens up the possibility that other
 decays could be significant. These can arise from 
$u\bar{u}$, $d\bar{d}$ components in the state, which can come
from mixing with a glueball.  

We solve Schwinger--type  
mass equations with linear masses, pioneered in refs. \cite{ten1,ten2}
and motivated in refs. \cite{ten2,sca,sch}. In this approach the underlying 
nature of the meson, whether conventional or hybrid, is not specified. 
The primitive (bare) states are ideally mixed.
Primitive isoscalar and isovector $u\bar{u}$, $d\bar{d}$
states are degenerate.
In this work we further only allow $SU(3)$ symmetric 
glueball--meson coupling, with no meson--meson coupling.
We restrict to ground state and first excited state mesons. It is known
that such restriction is quite accurate if the glueball mass is far from 
those of the states \cite{sch}, as is the case here.  

The numerical input is as follows. The ratio between pseudovector and
scalar glueball masses is evaluated in lattice QCD as
$1.70 \pm 0.05$ \cite{morn} or $1.73 \pm 0.09$ \cite{teper}. Taking the
world average scalar glueball mass as 1.6 GeV \cite{sca}, this implies
a (input) pseudovector glueball mass of 2.7 GeV. Our conclusions do not
critically depend on this value. The primitive $u\bar{u}+d\bar{d}$
ground state is input as the $b_1$ mass \cite{pdg98}. The physical
masses of $h_1(1170)$, $h_1^{'}(1380)$ and the newly discovered $h_1(1600)$ at
$1594\pm 15^{+10}_{-60}$ MeV \cite{e852} are used as input. We further 
assume that the difference between the primitive $s\bar{s}$ and
$u\bar{u}+d\bar{d}$ masses is the same for the ground states and 
excited states. Lastly, the primitive excited $u\bar{u}+d\bar{d}$ mass, 
the most uncertain input, is taken as $1650 \pm 50$ MeV. This is hence
the assumed mass region for the yet undiscovered excited $b_1$ resonance.

The output of our analysis is as follows. The 
experimenally unobserved physical excited $s\bar{s}$
state ($|h_1^{'}\rangle_2$) 
is predicted at $1810 \pm 40$ MeV. The difference between the
primitive $s\bar{s}$ and $u\bar{u}+d\bar{d}$ masses, for both the 
ground and excited states, is $180\pm 10$ MeV, yielding a primitive
$s\bar{s}$ ground state ($|s\bar{s}\rangle_1$) at $1410 \pm 10$ MeV. This is consistent with 
$1445\pm 41$ MeV derived from quark model 
relations\footnote{Combining $K(^1P_1)+K(^3P_1) = K(1270) + K(1400)$,
$\; b_1 + (s\bar{s})_1 = 2 K(^1P_1)$ and $K(^1P_1) - b_1 = K(^3P_1) - a_1$.
Here all items are the corresponding masses. The $^1P_1$ and $^3P_1$ kaon
masses before mixing are $K(^1P_1)$ and $K(^3P_1)$ respectively, and
the primitive $s\bar{s}$ ground state mass is $(s\bar{s})_1$.}. 
The coupling, in the notation of refs. \cite{ten1,ten2} is
$g_1 = 0.19 \pm 0.01$ GeV for the ground states and 
$g_2 = 0.19 \pm 0.12$ GeV for the excited states. The accurate former
value is larger than values found for scalar and tensor mesons 
\cite{ten1,sca}.

The valence content of the physical mesons is
$$|h_1\rangle _1\!=\!\left( \!-0.22^{+0.02}_{-0.01}\right) \!\!|g\rangle \!+\!
\left( 0.06^{+0.03}_{-0.05}\right) \!\!|s\bar{s}\rangle _2\!+\!
\left( 0.12^{+0.05}_{-0.09}\right) \!\!|u\bar{u}\rangle _2\!+\!
\left( 0.17^{+0.01}_{-0.00}\right) \!\!|s\bar{s}\rangle _1\!+\!
\left( {\bf 0.95}^{+0.01}_{-0.01}\right) \!\!|u\bar{u}\rangle _1,$$
$$|h_1'\rangle _1\!=\!\left( \!-0.13^{+0.02}_{-0.03}\right) \!\!|g\rangle \!+\!
\left( 0.06^{+0.03}_{-0.05}\right) \!\!|s\bar{s}\rangle _2\!+\!
\left( 0.13^{+0.08}_{-0.11}\right) \!\!|u\bar{u}\rangle _2\!+\!
\left( {\bf 0.96}^{+0.02}_{-0.03}\right) \!\!|s\bar{s}\rangle _1\!+\!
\left( \!-0.22^{+0.02}_{-0.03}\right) \!\!|u\bar{u}\rangle _1,$$
$$|h_1\rangle _2\!=\!\left( \!-0.19^{+0.15}_{-0.08}\right) \!\!|g\rangle \!+\!
\left( 0.16^{+0.09}_{-0.16}\right) \!\!|s\bar{s}\rangle _2\!+\!
\left( {\bf 0.94}^{+0.07}_{-0.08}\!\right) \!\!|u\bar{u}\rangle _2\!+\!
\left( \!-0.20^{+0.16}_{-0.11}\!\right) \!\!|s\bar{s}\rangle _1\!+\!
\left( \!-0.14^{+0.11}_{-0.06}\!\right) \!\!|u\bar{u}\rangle _1,$$
$$|h_1'\rangle _2\!\!=\!\!\left( \!-0.12^{+0.07}_{-0.01}\!\right) \!\!|g
\rangle \!+\!\left( {\bf 0.97}^{+0.04}_{-0.03}\!\right) \!\!|s\bar{s}\rangle _2\!+\!
\left( \!-0.21^{+0.22}_{-0.13}\!\right) \!\!|u\bar{u}\rangle _2\!+\!
\left( \!-0.06^{+0.03}_{-0.00}\!\right) \!\!|s\bar{s}\rangle _1\!+\!
\left( \!-0.05^{+0.03}_{-0.00}\!\right) \!\!|u\bar{u}\rangle _1.$$
where the states on the left and right are respectively the physical 
and primitive states. The first three physical states are the 
experimental states $h_1(1170)$, $h_1^{'}(1380)$ and 
$h_1(1600)$ \cite{pdg98}.

Decays are now studied by using a finite width for the initial meson, 
and unless otherwise indicated,
a narrow width approximation for the final mesons. Finite widths are
implemented by smearing over relativistic Breit--Wigner shapes with
Quigg -- von Hippel energy dependent widths.
Whenever a decay is OZI allowed from an ideally mixed initial state,
we assume, for simplicity,
that the initial state is $100\%$ ideally mixed. OZI
forbidden decays are implemented by using the (small) valence contents above
to calculate connected decays \cite{ten1}.

\begin{table}[tbh]
\caption{Partial decay widths of $h_1^{'}(1380)$ in MeV in relativistic
\protect\cite{biceps} and mock meson \protect\cite{kokoski87} 
phase space. The latter is in brackets. For conventional meson decays
in relativistic phase space
we allow the wave function parameter $\beta$, which is taken to be the
same for the incoming and outgoing states in a decay, to vary between
the reasonable values $0.35$ and $0.45$ GeV \protect\cite{biceps},
giving rise to the error estimate. The dagger indicates
that phase space is unreliable in the narrow resonance approximation for
the final state, so that the width is calculated by smearing over 
a Breit--Wigner for all broad resonances involved, 
both in the initial and final states. Since the $|u\bar{u}\rangle_2$
component of the physical $h_1^{'}(1380)$ has such a large uncertainty,
we only employ the $|u\bar{u}\rangle_1$ component for OZI forbidden 
decays. However, omission of the $|u\bar{u}\rangle_2$ component could 
significantly affect widths and especially $D/S$--wave width ratios.
 }
\label{table:1}
\newcommand{\m}{\hphantom{$-$}}
\newcommand{\cc}[1]{\multicolumn{1}{c}{#1}}
\renewcommand{\tabcolsep}{2pc} 
\renewcommand{\arraystretch}{1.2} 
\begin{tabular}{@{}lcc}
\hline
Mode & Wave   & Width \\
\hline
$K^{\ast}K\da$  & S & $137\pm 12$ \\
                & D & $1\pm 1$        \\
                &D/S& $0.010^{+0.008}_{-0.004}$        \\
$\rho\pi$       & S & $12\pm 3$    (13)  \\
                & D & $4\pm 3$    (4)  \\
                &D/S& $ 0.4^{+0.4}_{-0.2}$   (0.4)  \\
$\omega\eta$    & S & $2\pm 1$   (2)  \\
                & D &  0   (0)  \\
                &D/S& $ 0.01^{+0.01}_{-0.00}$   (0.01)  \\
$b_1\pi\da$     & P &  0      \\
\hline
Total                 &   &  156  \\
\hline
\end{tabular}\\[2pt]
\end{table}

The decays of conventional mesons are studied in the $^3P_0$ model using
the methods, conventions and parameters 
of refs. \cite{ten1,biceps}. 
Making the usual identification that the 
primitive ground state mesons are P--wave quark model states, we obtain 
the decay widths in Table 1. We note that although the 
experimentally observed $K^{\ast}K$ mode is dominant, and similar to the
total width of the state\footnote{We find that
mock meson phase space \protect\cite{kokoski87} 
gives a $K^{\ast}K$ partial width of
$191\pm 18$ MeV, inconsistent with the experimental total width of the
state \protect\cite{pdg98}. For near threshold decays of this type
mock meson phase space always gives a substantially larger width than
relativistic phase space. Mock meson phase space results are hence not
quoted for near threshold decays in the tables. We note that the 
$K^{\ast}K$ partial width calculation in ref. \protect\cite{kokoski87}
misses a flavour factor of two, and is hence unreliable.} \cite{pdg98}, 
the $\rho\pi$ mode
is detectable. It is not as large relative to $K^{\ast}K$
as one might expect from the limited phase space of $K^{\ast}K$.
Identification in $\rho\pi$ is complicated by the huge 
$360\pm 40$ MeV width of the $h_1(1170)$ mainly in $\rho\pi$ \cite{pdg98}.
This makes $\rho\pi$ an unattractive search channel for $h_1^{'}(1380)$,
since no viable production processes are known which strongly produce
the dominant $s\bar{s}$ component in $h_1^{'}(1380)$ as opposed to the 
dominant $u\bar{u}+d\bar{d}$ component in $h_1(1170)$.
Although $\omega\eta$ is small, $h_1^{'}(1380)$ has recently been
observed in this mode \cite{cbar}. Additional decay modes that have not
been calculated but are expected to be small
are decays to $h_1(\pi\pi)_S$ and direct three--body decays like
$\pi^0\pi^+\pi^-$.

\begin{table}[tbh]
\caption{Partial decay widths of pure $u\bar{u}+d\bar{d}$ 
$h_1(1594)$ (with the experimental total Breit--Wigner width $384$ MeV) 
in MeV for its interpretations as conventional and hybrid mesons.
Hybrid meson decays are calculated in the IKP and PSS models 
\protect\cite{pss}.
Other conventions are as in Table 1.
$h_1(1594)\rightarrow h_1(\pi\pi)_S$ is not estimated.
}
\label{table:2}
\newcommand{\m}{\hphantom{$-$}}
\newcommand{\cc}[1]{\multicolumn{1}{c}{#1}}
\renewcommand{\tabcolsep}{1.35pc} 
\renewcommand{\arraystretch}{1.2} 
\begin{tabular}{@{}lcccc}
\hline
Mode              & Wave & Meson & IKP Hybrid & PSS Hybrid \\
\hline
$\rho\pi$         & S &$14\pm 2$            (13)& 111 (96)& 86 (74)\\
                  & D &$126\pm 40$          (97)&   1  (1)&  1  (1)\\
                  &D/S&$9^{+1}_{-5}$       (7)& 0.005 (0.004)& 0.009 (0.008)\\
$\rho(1450)\pi\da$& S &$31\pm 1$                & 142     & 111    \\
                  & D &$6\pm 3$                 &   0     &   0    \\
                  &D/S&$0.2\pm 0.1$             & 0.0002  & 0.0004 \\
$K^{\ast}K$       & S &$15\pm 3$            (17)& 27  (31)& 37 (42)\\
                  & D &$17\pm 7$            (17)& 0    (0)& 0   (0)\\
            &D/S&$1.2^{+1.1}_{-0.6}$ (1.0)& 0.0004 (0.0003)& 0.0005 (0.0005)\\
$\omega\eta$      & S &$6\pm 2$              (6)& 19  (18)& 24 (23)\\
                  & D &$11\pm 5$            (10)& 0    (0)& 0   (0)\\
            &D/S&$1.8^{+1.8}_{-0.8}$ (1.6)& 0.002 (0.001)& 0.003 (0.002)\\
$b_1\pi$          & P & 0                       & 136 (227)& 0   (0)\\
\hline
Total             &   &225                      & 436     & 259    \\  
\hline
\end{tabular}\\[2pt]
\end{table}

We proceed to analyse $h_1(1600)$. One has to allow for the possibility that
the excited $u\bar{u}$, $d\bar{d}$
and $s\bar{s}$ states are hybrid mesons. The
calculations for this possibility are performed in the 
Isgur--Paton flux--tube
model with the standard parameters  of ref. \cite{pss}. The results are
displayed in Table 2. The $h_1(1600)$ is predicted to decay from most
to least prevalent to 
$\rho\pi\: / \:\rho(1450)\pi, \; K^{\ast}K$ and $\omega\eta$
in all interpretations of the state. 
A minor feature that distinguishes interpretations is the relative size
of the $\rho\pi$ and the $\rho(1450)\pi$ modes. 
The main distinguishing feature is the
ratio of $D$--wave to $S$--wave widths, which is consistently larger
for the meson than the hybrid interpretation. For the meson interpretation
the S--wave width is suppressed due to a node in the amplitude, making it
sensitive to the wave function parameter $\beta$ employed. Table
2 shows that the $D/S$--wave width ratio in $\omega\eta$ 
is inconsistent with the
experimental result $0.3^{+0.1}_{-0.[1-3]}$ \cite{priv} if $h_1(1600)$
is a conventional meson. In order to confirm this result, we perform three
further checks. Firstly, we evaluate the ratio by taking the wave function
parameter $\beta$ to be different for different
mesons participating in the decay. Varying $\beta$ in the
reasonable range $0.35 - 0.45$ GeV \cite{biceps} confirms the result.
Secondly, using the full valence content of $h_1(1600)$ above, and
allowing decay via the ground state P--wave meson component, 
confirms the result. Thirdly, experimental data has few D--wave events
above 1.8 GeV \cite{e852}. Restricting the Breit--Wigner smearing to 
invariant masses less than 1.8 GeV gives the nearest ratio to experiment
in all these simulations, $0.9^{+1.2}_{-0.5}$. This ratio is still
outside the range allowed by experiment, although it is not far outside the
range. Experimentally, it has not been
established that the D--wave exists \cite{priv}, so that the
very small ratio predicted for a hybrid meson in Table 2 could be
consistent with experiment.
Thus current experiment makes the conventional meson interpretation of the
$h_1(1600)$ unlikely, but allows the hybrid interpretation. 
This assumes that the state observed 
in experiment cannot be resolved into two separate states. 
Since the 
$^3P_0$ model has only been tested for a few $D/S$--wave width ratios
\cite{biceps}, one needs further information.
The total width 
$384\pm 60^{+70}_{-100}$ MeV of $h_1(1600)$ is slightly
more consistent with the hybrid
interpretation.
Future searches for $h_1(1600)$ should
focus on obtaining the $D/S$--wave width ratio in the
sizable $\rho\pi$ channel.
The $b_1\pi$ mode distinguishes
the two models of hybrid decay in Table 2.

We note that since the $\rho$ Regge trajectory dominates the
$\rho(1450)$ and $b_1$ trajectories, and $h_1(1600)$ has a healthy
$\rho\pi$ coupling for all interpretations,
one expects the $h_1(1600)$
to be produced via natural parity exchange in the $\pi^- p$ collisions
it has been observed in. This is confirmed in the experimental analysis
\cite{e852}, providing an independent check on our calculations.
The non--observation of $h_1(1600)$ in unnatural parity exchange
\cite{e852} 
may put bounds on its $b_1\pi$ coupling, discriminating between
different hybrid decay models.

\begin{table}[tbh]
\caption{Partial decay widths of $b_1(1650)$ in MeV. 
Conventions are as in Table 2, including using the same total width
for $b_1(1650)$ as for $h_1(1594)$.
$b_1(1650)\rightarrow b_1(\pi\pi)_S$ is not estimated.
}
\label{table:3}
\newcommand{\m}{\hphantom{$-$}}
\newcommand{\cc}[1]{\multicolumn{1}{c}{#1}}
\renewcommand{\tabcolsep}{1.28pc} 
\renewcommand{\arraystretch}{1.2} 
\begin{tabular}{@{}lcccc}
\hline
Mode              & Wave & Meson & IKP Hybrid & PSS Hybrid \\
\hline
$\omega\pi$       & S &$4^{+2}_{-0}$             (4)& 37  (30)& 28 (22)\\
                  & D &$48\pm 13$               (35)&  0 (0)& 0 (0)\\
                  &D/S&$11^{+0}_{-3}$       (9)& 0.006 (0.005)& 0.01 (0.01)\\
$\omega(1420)\pi\da$&S&$11\pm 1$                    & 70    & 54   \\
                  & D &$7\pm 3$                     &  0    &  0  \\
                  &D/S&$0.6^{+0.6}_{-0.2}$          &0.0009&0.001\\
$K^{\ast}K$       & S &$13\pm 3$                (14)& 30  (32)& 40 (42)\\
                  & D &$23\pm 9$                (22)&  0 (0)& 0 (0)\\
               &D/S&$1.8^{+1.7}_{-0.9}$ (1.5)&0.0005 (0.0004)&0.0007(0.0007)\\
$\rho\rho\da$     & S &$34\pm6$                     &  0 & 0 \\
                  & D &$34\pm 15$                   &  0 & 0 \\
                  &D/S&$1.0^{+0.8}_{-0.4}$          &     &    \\
$\rho\eta$        & S &$5\pm 1$                  (5)& 20  (18)& 25 (22)\\
                  & D &$15\pm 6$                (12)&  0  (0)& 0 (0)\\
                &D/S&$3.1^{+2.6}_{-1.6}$ (2.6)&0.002 (0.002)& 0.003 (0.003)\\
$a_0\pi\da$       & P &$8\pm 1$                     & 56    & 3   \\
$a_1\pi$          & P &$11\pm 2$                (16)& 19  (30)& 3 (5)\\
$a_2\pi$          & P &$82\pm 16$              (132)& 37  (60)& 7 (12)\\
                  & F &$3^{+4}_{-2}$             (4)&  0 (0)& 0 (0)\\
           &F/P&$0.03^{+0.04}_{-0.01}$ (0.03)&0.005 (0.005)&0.0003 (0.0003)\\
$h_1\pi$          & P & 0                           & 72  (108)& 0 (0)\\
\hline
Total             &   & 296                         & 341&160\\  
\hline
\end{tabular}\\[2pt]
\end{table}

In Table 3 the widths for the isopartner state $b_1(1650)$ are calculated.
The channels that distinguish between conventional and hybrid meson
interpretations of the state, $\omega(1420)\pi$ and $\rho\rho$, are
difficult to access experimentally. However, $D/S$--wave width ratios
remain an excellent distinguishing feature. Possible search channels
are $\omega\pi$ and $\rho\eta$.

\begin{table}[tbh]
\caption{Partial decay widths of pure $s\bar{s}$ $h_1^{'}(1810)$ in MeV.
Conventions are as in Table 2, except that $h_1^{'}$ has a total width
of 100 MeV. $h_1^{'}(1810)\rightarrow h_1^{'}(1380)(\pi\pi)_S$ is not estimated.}
\label{table:4}
\newcommand{\m}{\hphantom{$-$}}
\newcommand{\cc}[1]{\multicolumn{1}{c}{#1}}
\renewcommand{\tabcolsep}{1.28pc} 
\renewcommand{\arraystretch}{1.2} 
\begin{tabular}{@{}lcccc}
\hline
Mode              & Wave & Meson & IKP Hybrid & PSS Hybrid \\
\hline
$K^{\ast}K$     & S &$11\pm 9$  (10)& 47  (43)& 47 (43)\\
                & D &$70\pm 30$  (61)&  0 (0)& 0 (0)\\
                &D/S&$6^{+22}_{-4}$  (6)& 0.004  (0.004)& 0.009 (0.008)\\
$\phi\eta$      & S &$17\pm 5$  (18)& 22  (24)& 56 (60)\\
                & D &$14\pm 7$  (14)&  0 (0)& 0 (0)\\
                &D/S&$0.8^{+1.3}_{-0.4}$  (0.8)& 0.0004  (0.0004)& 0.0006 (0.0006)\\
$K_1(1270)K\da$ & P &$ 1\pm 0$   &  13   & 0   \\
\hline
Total                   &   & 113& 82& 103\\  
\hline
\end{tabular}\\[2pt]
\end{table}

The widths for the undiscovered excited $s\bar{s}$ state 
$h_1^{'}(1810)$ are indicated in
Table 4. 
It is interesting to note that the flux--tube model selection rule,
which states that decays to $S+S$ states ($K^{\ast}K,\; \phi\eta$) are 
suppressed relative to $P+S$ states ($K_1(1270)K$) \cite{pss},
is apparently violated. This is due to phase space. 
Whether the $h_1^{'}(1810)$  is a conventional or 
hybrid meson, it is surprisingly narrow.
Excellent search channels are $K^{\ast}K$ and $\phi\eta$. The latter
is especially interesting since it cannot come from a $u\bar{u}+d\bar{d}$
state via OZI allowed decay. Small OZI 
forbidden modes like $\rho\pi$ could also effect detection.
A natural place to search for $s\bar{s}$ states is at Jefferson Lab,
where the photon has a sizable coupling to $s\bar{s}$. Production is
likely to be via diffractive exchange, as meson exchange involves
OZI forbidden or evading processes.

\vspace{.2cm}

We thank C. Amsler, P. Eugenio, E. Klempt
 and D. Weygand for useful discussions on 
their experimental data. This research is supported by the Department of 
Energy under contract W-7405-ENG-36.

\end{document}